\documentclass{article}
\usepackage{amsthm,amsmath}
\usepackage{tikz,pgf}
\usepackage{nicematrix}
\usetikzlibrary{shapes.geometric}

\usepackage{graphicx}
\usepackage{cite}

\newtheorem{theorem}{Theorem}[section]

\title{Structural Analysis of Multi-core Processor and Reliability Evaluation Model}

\date{}

\author{
  Sergo Tsiramua\footnote{University of Georgia, Tbilisi, Georgia,
  s.tsiramua@ug.edu.ge}\and
  Hamlet Meladze\footnote{Muskhelishvili Institute of Computational Mathematics, Georgian Technical University, Tbilisi, Georgia,
  h\_meladze@hotmail.com}\and
  Tinatin Davitashvili\footnote{Ivane Javakhishvili Tbilisi State University, Tbilisi, Georgia, tinatin.davitashvili@tsu.ge}\and
  J. M. Sanchez\footnote{Faculty of Education, University of Malaga}\and
  F. Criado-Aldeanueva\footnote{¶Department of Applied Physics II Polytechnic School, Malaga University} }

\begin{document}
\maketitle

\begin{abstract}
In the present paper the models of structural analysis and evaluation of efficiency indicators (reliability, fault tolerance, viability, flexibility) of a multi-core processor with variable structure, equipped with multi-functional cores, are considered.

Using logical-probabilistic methods, the following has been developed: Models for evaluating the reliability and fault-tolerance of processor cores as multi- functional elements; Logical-probabilistic models of the shortest paths, flexibility and performance conditions for successful operation of multi-core processors based on multi-functional cores; Models for estimating reliability, fault tolerance and lifetime of multi-core processors considering all possible states of performance.

The results of the structural analysis of two-core and four-core processors and the trends of increasing the efficiency indicators of multi-core processors are presented.
  \\

\noindent\textbf{MSC2020-Mathematics Subject Classification System:} 68M15 Reliability, testing and fault tolerance of networks and computer systems; 93B12 Variable structure systems.
\\

\end{abstract}

\section{Introduction}
\label{sec:1}

Currently, a fundamentally new methodology for scientific research has been formulated: mathematical modeling and computational experiment on modern computers. Using a computational experiment, large and complex problems that require huge calculations are solved. Such high
performance can be achieved based on the development of parallel computational algorithms (see e.g. \cite{1, 2, 3, 4, 5, 6, 7, 8}, where the algorithms for parallel computing systems are developed). Along with this, there arises a problem with the reliability of supercomputers.

Multi-core processors are widely used in modern computers to increase
performance, reliability and energy efficiency. A multi-core processor
is the tightly connected cores, integrated into a single crystal with
a common cache memory. Multi-core processors are found in a variety of
computing systems, including central processing units (CPUs) of
general-purpose personal computers, networking, embedded, digital
signal processing (DSP), computer graphics (GPU), mobile phones, and
other devices. In multi-core processors, within one cycle, each core
goes through four main steps, namely: receiving instructions and data,
decoding instructions (for example, where the current instruction ends
and where the next one begins), execution (for example, performing
arithmetic or logical operations in an arithmetic-logic unit) and
memorizing the received result in processor registers for quick
access, which is eventually transferred to RAM.

The number of cores in general-purpose personal computer's processors
is measured in the tens, the number of cores for specialized chips can
exceed $10.000$, and the number of cores in supercomputers (i.e.,
clusters of chips) can exceed millions. Accordingly, structural
analysis of multi-core processor systems, quantitative evaluation and
optimization of reliability and efficiency indicators due to the
substantial number of all possible states of the system, belong to the
class of large-scale, complex problems.

A common feature of multi-core processors is their ability to function
in parallel mode, which ensures:
\begin{enumerate}
\item performance improvement -- execution of the algorithm of a problem in parallel mode reduces the time of execution;
\item increasing economy -- in a parallel system, the operating system distributes the execution of the problem between the cores;
\item energy efficiency -- by distributing tasks between cores, it is possible to scale energy consumption;
\item increasing reliability -- when one of the cores fails, the operating system replaces this core with another core;
\item improving fault tolerance -- in case of hardware or software faults in any of the cores, the multi-core system will continue to function successfully.
\end{enumerate}

In multi-core processors, the reliability of the cores can be affected
by many factors, one of which is power consumption and temperature
changes in the cores at different clock frequencies, which negatively
affects the reliability of the processor. To solve the problem, in
order to achieve a balance between energy consumption and reliability,
dynamic control of power supply, considering reliability is proposed
in \cite{9}. In multi-core processors, there are many possibilities to
scale the energy consumption if tasks are distributed among the
cores. As the authors of \cite{9} argue, the parallelism of the
functioning of multi-core processors should be used not only to
increase performance but also to increase reliability.

\cite{10} discusses the failures of the temporary or permanent nature
of individual elements in multi-core processors. A temporary failure
can happen once and not be repeated again. A permanent failure is
mainly due to wear of the device and has a permanent nature. The
article proposes dynamic tools for checking the integrity of computing
processes and a method of increasing reliability at the
microarchitectural level.

In \cite{11}, in order to increase the performance and reliability of
multi-core processors, the creation of a new computing environment is
considered, in which, in case of failure of a core's functional block
(computing element), its function is transferred to a private or
shared memory block. This process can be performed when a functional
block experiences a failure (moves into an idle state), while a
private or shared memory block maintains a functional state. With such
an approach, it is obvious that the fault tolerance and the viability
of the multi-core processor are increased.

\cite{12} discusses the reliability issues of heterogeneous multi-core
processors. In the article is shown that even minor errors in problem
processing by non-identical cores are a significant problem. To solve
the problem, in the article is given the detailed investigation of
each running program's reliability characteristics with further
dynamically distribution of tasks between different types of cores.

\cite{13} discusses NASA's stringent reliability requirements for
cosmos multi-core computing in terms of power, mass, and cost
(expenses). If the processor core, switches, I/O or memory ports fail,
the computing load must be transferred to other resources on the
circuit. Such an approach is not perfect, since there are some
failures that can bring the entire chip out of order. But in general,
the proposed approach can lead to the significant increase in overall
reliability, that is, the period of time during which the computer can
ensure the performance of the task. The effectiveness of this approach
depends on a design of multi-core chip that is optimized to minimize
the probability of a single chip failure, and an architecture that can
effectively eliminate the failure of cores, memory, and I/O elements.

The authors of article \cite{14} fully share the opinions and
approaches of the authors of discussed articles. As in homogeneous
multi-core processors, which include only identical cores, in
heterogeneous multi-core processors, which consist of non-identical
cores, cores with corresponding blocks are considered as
multi-functional elements of a multi-core processor system, since any
core can perform different types of tasks, and to increase reliability
tasks can be distributed among cores.  Considering full or partial
core failures, multi-core processors belong to the class of
multi-state systems, whose reliability models more realistically
describe the technical systems, as evidenced by the review of studies
in the article \cite{14}.

The work in \cite{15} is likely the most closely related to the approach of enhancing the reliability of multi-core processors by utilizing their structural redundancy. According to the authors, the widespread use of multi-core processors is due to their flexibility, high performance, low power consumption, and internal redundancy. Despite the advantages of multi-core processors, there are some reliability issues that must be addressed to make them suitable for critical tasks. The reliability and safety of systems based on these devices are reduced due to the increasing complexity and scale of integration. The authors propose an approach based on N-modular redundancy (NMR) and M-partitioning to enhance the reliability of parallel applications running on multi-core processors. To validate the proposal, a case study was conducted where two parallel benchmark applications ran using additional resources on the chip. The efficiency of this approach was evaluated using the software-implemented fault injection (SWIFI) method.

As technology scales, the power density of multi-core chips increases, leading to the formation of hotspots that accelerate device aging and cause chip failure. Intense efforts to reduce power consumption by using low-power methods reduce the reliability of new generations of processors. To evenly distribute temperature across all cores, a proactive mechanism is needed to predict future workload characteristics and corresponding temperatures, enabling decisions to be made before hotspots occur. These proactive methods rely on an engine for accurate prediction of future workload characteristics. In \cite{16}, the authors discuss modern methods for predicting workload dynamics and compare their performance. They present a forecasting method based on support vector regression (SVR), which accurately predicts the behavior of workloads several steps ahead. To assess the effectiveness of the approach, the authors use the programs from the set of tests PARSEC on the UltraSPARC T1 processor running the Sun Solaris operating system and extract architectural traces. The extracted traces are then used to generate power and temperature profiles for each core using the McPAT and Hot-Spot simulators. The results show that the proposed method accurately predicts workload and power dynamics, outperforming previous forecasting methods.

In \cite{17}, a new approach is presented that uses machine learning to improve CPU scheduling in heterogeneous multi-core systems by developing a matching methodology that increases system throughput. The proposed scheduling approach uses Long Short-Term Memory (LSTM), artificial neural networks, and linear regression to predict parameters such as instructions per cycle for threads running on different cores. The predicted values are then used for scheduling threads using matching, which maximizes the number of instructions per cycle. The results show that the artificial neural network outperforms linear regression in predicting instructions per cycle. According to the authors, compared to traditional heterogeneous scheduling, the proposed machine learning-based scheduling approach improves the performance of the heterogeneous system by approximately 1.2 times and increases throughput by 20\%.

The high performance of the underlying frameworks enables the generation and collection of more data related to errors/failures, considering complex software stack configurations, in a reasonable amount of time. When dealing with large datasets related to multicore CPU failures obtained from multiple failure campaigns, it is important to filter out parameters (i.e., functions) that have no direct correlation with the system analysis of soft errors. In \cite{18}, controlled and uncontrolled machine learning methods are proposed to eliminate irrelevant information and detect correlations between failure injection results and application and platform characteristics. This new approach provides the necessary tools to explore new and more effective failure mitigation methods.

Regarding the reliability of multi-core processors, the articles discussed here and generally published in scientific journals rarely address the analysis of the structural/functional redundancy of multi-core processors and individual core execution units, as well as the issues of quantitative reliability assessment modeling from this perspective.

In this paper, we present logical-probabilistic methods for studying the structural/functional redundancy of multi-core systems and the results of combinatorial analysis. For structural analysis, we initially construct a logical matrix and logical function of the system’s functional resources (0, 1), describing the reconfigurable multi-core system’s structure. With the help of this matrix and logical function, we investigate the system’s structure and flexibility, determine the shortest operational paths, and evaluate the conditions for its functionality. At this stage, we also determine the total number of possible system states and the proportion of those states that represent operational conditions.

By transforming the logical functions of the multi-core system’s structure using logical-pro\-ba\-bi\-lis\-tic methods, we obtain a form of the logical function where logical elements can be replaced with probabilistic data, and logical operations are substituted by mathematical operations. This approach allows us to derive a probabilistic representation of estimation of the system's reliability and perform a probabilistic assessment of reliability.

In the reliability assessment model for multi-core systems, the initial data consist not of physical elements (such as cores or blocks), but of their functional capabilities, represented as logical or probabilistic variables. These probabilistic data can be derived, based on statistical data and using Fuzzy Logic, or the Monte Carlo method. Using these data, we construct a probabilistic matrix of the system's functional capabilities, which describes the potential capabilities of the multi-core system.

In the reliability model of multi-core systems, we consider the following constraints: the system functions in parallel mode, with independent and non-recoverable partial failures. Additionally, in the reliability model of a multi-core system, in the case of partial failure of components, we account for an instantaneous switch to an operational mode that ensures the system's fault tolerance and the continuation of successful functionality. While it is assumed that the reliability model includes relatively strict constraints, the combinatorial and logical-probabilistic analysis of their structure, in a static mode, enables us to determine the potential capabilities of the system and perform a comparative analysis of various models \cite{15, 16, 19, 20, 21}.

In future research, we will explore the dynamic processes of multi-core system operation, which can be described using a Markov model, taking time-related factors into account.

\section{The reliability assessment model of a multi-functional \\ element}
\label{sec:2}

A \textit{multifunctional element} (MFE) is an element with functional redundancy, which has the ability from the $m > 1$ number of functions assigned to the system, to perform any function $f$ at any moment $t$, from the set of its functional resources (functional capabilities) $F_a = \{f_e |\; e \in [1,k]\}$, $k>1$ and, if necessary, switch to performing another function from the set $F_a$ \cite{15, 16, 19, 20, 21}.

In multi-core processors, each core at any time moment can execute a certain instruction (algorithm, assignment, task) of the program (application), which can be called a function assigned to the core at a given moment of time. Thus, in an $n$-core processor $A=\{a_1,a_2,\dots,a_n\}$ the task $F=\{f_1,f_2,\dots,f_n\}$ is distributed among the cores that perform these tasks in sequence, parallel or in a mixed (sequence-parallel) mode.

The number of cores $n$ in the processor, the number of algorithms (tasks, functions) $m$ to be distributed between the cores and the number of functional resources $k_i$ of each core of the processor represent the structural parameters of the multi-core processor. The value of the system efficiency indicators depends on the size of these parameters, their ratio ($n > m > k_i$) and the structure of the function distribution between the cores, which will be shown
below. Depending on the number $m$ of functions of the set $F$, imposed on the multi-core processor, the MFE for a given system can be functionally complete $(k = m$, for example, in homogeneous multi-core processors) or functionally incomplete $(k < m$, for example, in heterogeneous multi-core processors due to non-identical cores).

A system composed of multifunctional elements has many ways of operation that can be switched to in case of failure of any element. For example, in functioning a multi-core processor, it might
happen partial or entire failure of cores as multi-functional elements with respect to the performance of the task algorithm. In the case of complete failure of the processor core, the core will be replaced with another working core, and in the case of partial failure of the core,
the operating system can redistribute the task execution algorithm between the cores (i.e., implement interchangeability the cores), which ensures the successful execution of the task by the processor. Thus, a multi-core processor is an adaptable system with a reconfigurable structure, which has the property of maneuvering (reconfiguring) to ensure successful operation.

Each core of a multi-core processor represents a multifunctional element (multifunctional element, MFE). When making quantitative assessment of its reliability, it should be considered the ability to perform any one function at any moment of time or not to perform one, two, etc. functions from the set of its functional resources (functional capabilities) $F_a=\{f_e |\; e\in[1,k]\}$, $k>1$.

In the article, we use $F_A=\{f_1,f_2,\dots,f_m\}$, $i\in[1,m]$, to denote the number of tasks assigned to system $A$ (a multicore processor) at time moment $t$, or the number of functions assigned to the core execution blocks. We use $F_{a_i}=\{f_1,f_2,\dots,f_{ki}\}$, $1<k_i\leq m$, to represent the functional capacity of each core (or block), with $F_{a_i}\subseteq F_A$. When $k_i=m$, we refer to the $i$-th element as a functionally full multifunctional element. When $k_i<m$, the $i$-th element is referred to as a functionally incomplete element.

Operability of MFE is such a state of the element, when its main parameters are within the acceptable limits, which correspond to the ability of the element to perform at least one function from the set of its functional resources.

Inoperability of the MFE is its condition when it cannot perform any of its functions from the set of functional resources.

A complete failure of MFE is an event when an element goes from an operable state to a non-operable state.

Partial failure of MFE is an event when the element's loss of ability to perform one or more functions does not lead transition to the state of inoperability and the element remains in the state of operability.

The MFE's reliability assessment model should include all the operability states in which it might be appear during functioning. When partial failures occur during the operating of MFE,
the number of its functional capabilities is reduced, and when it completely exhausts the functional capability resource, a full failure state occurs (although the MFE can go from a functionally full state to a state of full failure instantaneously without going through the
partial failure phase).

\begin{figure}[htbp]
  \center
  \begin{tikzpicture}
    \draw (-1,0) -- (0,0);
    \draw (3,0) -- (4,0);
    \draw (0,-1.5) -- (0,1.5);
    \draw (3,-1.5) -- (3,1.5);
    \foreach \y/\l in {-1.5/$f_k$,.5/$f_2$,1.5/$f_1$} {
      \draw (0,\y) -- (3,\y);
      \draw  (1.5,\y) node[rectangle,fill=white,text opacity=1,fill opacity=1,draw] {\l};
    }
    \draw (1.5,-.5) node {$\vdots$};
    \draw (.9,-1.9) rectangle ++(1.2,3.8);
    \draw (1.5,2) node {$a$};
    \draw (3.5,0) node[anchor=south] {$F_a$};
  \end{tikzpicture}
  \caption{MFE reliability model}\label{fig:1}
\end{figure}
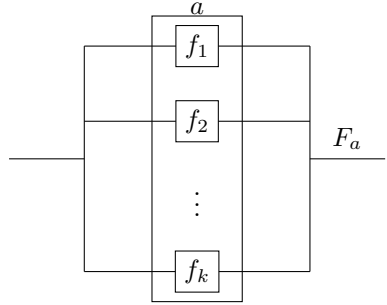

As an example, we consider the $k$-functional schematic model of reliability of MFE, the set of functional resources of which is $F_{a_i} = \{f_e |\; e\in[1,k_i]\}$ (see Figure \ref{fig:1}).

The set of all possible states of MFE
\begin{equation*}
  G = \Big\{
  (f_1,f_2,\dots,f_k),
  (f_1',f_2,\dots,f_k),
  \ldots,
  (f_1',f'_2,\dots,f_{k-1}',f_k),
  (f_1',f'_2,\dots,f_k')
  \Big\}
\end{equation*}
is a set of such states when at any time moment $t$ it is possible to have the ability to perform all functions (be in a fully functional state), or be in a state of partial failure or full failure. The number of all possible states of the $k$-functional MFE is $d=\text{card}\{G\}=2^k$, from which the number of operability states will be $d^*=\text{card}\{G^*\}=2^k-1$.

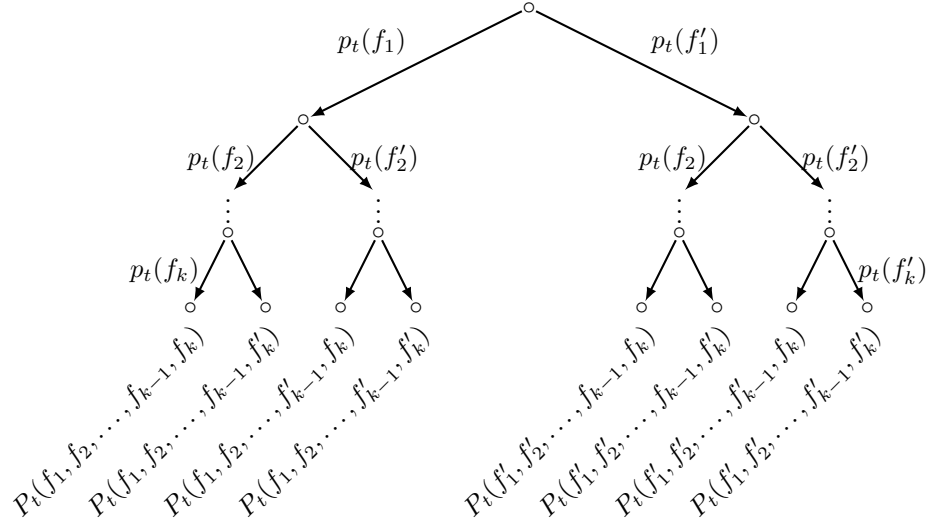
\begin{figure}[htbp]
  \center
  \begin{tikzpicture}
    \draw (-5,0) node {$\circ$} node[anchor=north east,rotate=45] {$P_t(f_1,f_2,\dots,f_{k-1},f_k)$};
    \draw (-4,0) node {$\circ$} node[anchor=north east,rotate=45] {$P_t(f_1,f_2,\dots,f_{k-1},f_k')$};
    \draw (-3,0) node {$\circ$} node[anchor=north east,rotate=45] {$P_t(f_1,f_2,\dots,f'_{k-1},f_k)$};
    \draw (-2,0) node {$\circ$} node[anchor=north east,rotate=45] {$P_t(f_1,f_2,\dots,f'_{k-1},f_k')$};
    \draw (1,0) node {$\circ$} node[anchor=north east,rotate=45] {$P_t(f_1',f_2',\dots,f_{k-1},f_k)$};
    \draw (2,0) node {$\circ$} node[anchor=north east,rotate=45] {$P_t(f_1',f_2',\dots,f_{k-1},f_k')$};
    \draw (3,0) node {$\circ$} node[anchor=north east,rotate=45] {$P_t(f_1',f_2',\dots,f'_{k-1},f_k)$};
    \draw (4,0) node {$\circ$} node[anchor=north east,rotate=45] {$P_t(f_1',f_2',\dots,f'_{k-1},f_k')$};

    \foreach \x/\y in {-2.5/1,1.5/1} {
      \draw [-latex,thick,shorten >=3pt,shorten <=3pt]
      (\x,\y) node {$\circ$} node[anchor=south] {$\vdots$} -- (\x-.5,\y-1);
      \draw [-latex,thick,shorten >=3pt,shorten <=3pt] (\x,\y) -- (\x+.5,\y-1);
    }

    \draw [-latex,thick,shorten >=3pt,shorten <=3pt]
    (-4.5,1) node {$\circ$} node[anchor=south] {$\vdots$}
    -- node[pos=.5,anchor=east] {$p_t(f_k)$} (-5,0);
    \draw [-latex,thick,shorten >=3pt,shorten <=3pt] (-4.5,1) -- (-4,0);

    \draw [-latex,thick,shorten >=3pt,shorten <=3pt]
    (3.5,1) node {$\circ$} node[anchor=south] {$\vdots$} -- (3,0);
    \draw [-latex,thick,shorten >=3pt,shorten <=3pt] (3.5,1) --node[pos=.5,anchor=west] {$p_t(f'_k)$} (4,0);

    \foreach \x/\y in {-3.5/2.5,2.5/2.5} {
      \draw [-latex,thick,shorten >=3pt,shorten <=3pt]
      (\x,\y) node {$\circ$}
      -- node[pos=.5,anchor=east] {$p_t(f_2)$} (\x-1,\y-1);
      \draw [-latex,thick,shorten >=3pt,shorten <=3pt] (\x,\y) -- node[pos=.5,anchor=west] {$p_t(f'_2)$} (\x+1,\y-1);
    }
    \foreach \x/\y in {-.5/4} {
      \draw [-latex,thick,shorten >=3pt,shorten <=3pt]
      (\x,\y) node {$\circ$}
      -- node[pos=.5,anchor=south east] {$p_t(f_1)$} (\x-3,\y-1.5);
      \draw [-latex,thick,shorten >=3pt,shorten <=3pt] (\x,\y) -- node[pos=.5,anchor=south west] {$p_t(f'_1)$} (\x+3,\y-1.5);
    }
  \end{tikzpicture}
  \caption{Tree of all possible states of MFE}\label{fig:2}
\end{figure}

The tree of all possible states of MFE are shown on Figure \ref{fig:2}, where all branches correspond to the MFE's operational state except the rightmost branch, when the MFE has lost the ability to perform all functions and is in a complete failure state.

The probability that the MFE has the ability to perform all the
functions assigned to the system, and at any time $t$ performs anyone
function of the set $F_a$ is expressed by the formula
\begin{equation}
  P_t(f_1,f_2,\dots,f_k) = \prod_{j=1}^k p_t(f_j),\label{eq:1}
\end{equation}
where $p_t(f_j)$ is the probability that function $f_j$, $j\in[1,k]$,
will be executed by the element at a time moment $t$.

The probability that at a time moment $t$ the MFE can perform any one
function of the set $F_a$ except the $k$-th function will be:
\begin{equation}
  P_t(f_1,f_2,\dots,f'_k)
  = \prod_{j=1}^{k-1} p_t(f_j) \cdot (1 - p_t(f_k))
  = \prod_{j=1}^{k-1} p_t(f_j) q_t(f_k)
  \label{eq:2}
\end{equation}

The probability that the MFE cannot perform any of the functions from
the set $F_a$ at a time moment $t$ will be:
\begin{equation}
  P_t(f'_1,f'_2,\dots,f'_k)
  = \prod_{j=1}^{k} (1 - p_t(f_j))
  = \prod_{j=1}^{k} q_t(f_j)
  \label{eq:3}
\end{equation}

In the given formulas, $p_t(f_j)$ is a probability of correct (without
any failure) performance of the $j$-th function of the $F_a$ set
at time $t$, and $q_t(f_j)$, is a probability of failure with respect
to the function $f_j$.

It is clear that according to the complete probability theorem, the
sum of the probabilities of all states of the MFE is equal to
1. Therefore, the reliability of MFE -- the probability that at a given
time $t$ the element has an ability to perform at least one function
from the $F_a$ set, i.e. is in a functional state, will be:
\begin{equation}
  P_a(t) = 1 - \prod_{j=1}^{k} (1 - p_t(f_j))
  \label{eq:4}
\end{equation}
where $P_a(t)$ is a probability that MFE can operate at a time moment
$t$, and $p_t(f_j)$, $j\in[1,k]$, is a probability that MFE can
perform each function \cite{22}.

\section{Reliability of structurally reconfigurable systems}
\label{sec:3}

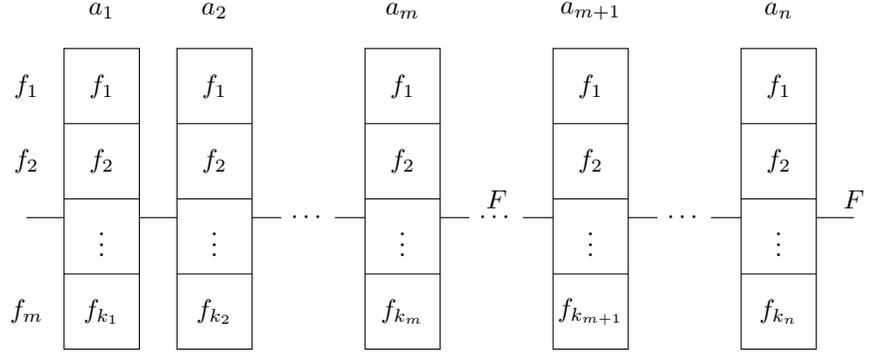
\begin{figure}[htbp]
  \center
  \begin{tikzpicture}
    \draw (-1,.25) -- (10,.25);
    \foreach \x/\l/\s in {0/$f_{k_1}$/$a_1$,1.5/$f_{k_2}$/$a_2$,4/$f_{k_m}$/$a_m$,6.5/$f_{k_{m+1}}$/$a_{m+1}$,9/$f_{k_n}$/$a_n$} {
      \draw[fill=white] (\x-.5,-1.5) rectangle (\x+.5,2.5);
      \foreach \y in {-.5,.5,1.5} {
        \draw (\x-.5,\y) -- (\x+.5,\y);
      }
      \draw (\x,3) node {\s};
      \draw (\x,2) node {$f_1$};
      \draw (\x,1) node {$f_2$};
      \draw (\x,0) node{$\vdots$};
      \draw (\x,-1) node {\l};
    }
    \draw[fill=white] (2.75,.25) node[rectangle,fill=white,text opacity=1,fill opacity=1] {$\cdots$};
    \draw[fill=white] (5.25,.25) node[rectangle,fill=white,text opacity=1,fill opacity=1] {$\cdots$} node[anchor=south] {$F$};
    \draw[fill=white] (7.75,.25) node[rectangle,fill=white,text opacity=1,fill opacity=1] {$\cdots$};
    \foreach \y/\l in {-1/$f_m$,1/$f_2$,2/$f_1$} {
      \draw (-1,\y) node {\l};
    }
    \draw (10,.25) node[anchor=south] {$F$};
  \end{tikzpicture}
  \caption{Reliability model of the MFE-based assembled system}\label{fig:3}
\end{figure}

The general schematic model of the reliability of the variable
structure system composed of MFEs can be presented in the following
form (see Figure \ref{fig:3}).

In case of partial failure of any element in the system composed on the basis of MFEs, it is possible to redistribute the functions between the elements, which ensures the successful operation of the system. Such a possibility makes the system flexible. Accordingly, a multi-core processor based on multi-functional cores is a structurally reconfigurable system.

During system operation in a parallel mode, in case of partial failure of one of the MFEs, in order to continue successful operation, it is possible to interchange the elements, and in case of complete failure of the MFE, it is necessary to turn on the backup element. In the mode
of execution of functions in order, in addition to mutual replacement, it is also possible to replace elements. That is, after performing its function, the MFE can replace another element that cannot perform the function due to it at that moment.

The structure of the system constructed on the basis of MFE is determined by the composition of functional elements, their interconnections and the scheme of distribution of functions between
elements. As mentioned above, the main parameters of the structure of reconfigurable systems are:
\begin{itemize}
\item[$n$:] the number of cores of a multi-core processor system $A=\{a_i\}$, $i\in [1,n]$;

\item[$m$:] the number of functions (tasks, assignments) $F=(f_1,f_2,\dots,f_m)$ imposed on the system $A$;

\item[$k_i$:] the number of functional resources of each $i$-th MFE   core;

\item[$k_\Sigma$:] the sum of functional resources of all MFEs (number of functional resources of the system);

\item[$\delta_S$:] the scheme of distribution of functions between elements (structure).
\end{itemize}

Performance indicators of the system -- the characteristics of reliability, flexibility, resistance to failure, vitality are uniquely dependent on the specified structural parameters and their interrelationship.

As mentioned above, a multi-core processor system has a flexible structure and is reconfigurable due to the functional redundancy of cores. If a partial failure of any core occurs in relation to the assigned function, it will be possible such replacement of the elements (rearrange the structure of the system) when the condition of simultaneous performance of all the functions assigned to the system is restored. It should also be noted here that the effective functioning of such systems depends not only on $n$, $m$, $k$ and $k_\Sigma$ parameters, but also on the scheme $\delta_S$ of distribution of functions between MFEs.

In order to evaluate the system's reliability (infallible operation and resistance to failure), we describe the condition of its operability.

By the system $A=\{a_i|\;i\in[1,n]\}$ the function $F =\{f_j|\;k\in[1,m]\}$ is executed successfully if at a given time moment all $n^*=m$ $(n^*\le n)$ elements complete all $f_j$, $j\in[1,m]$
functions from the set $F$, imposed on the system, so that each element at any time moment performs only one defined function from the set of its functional resources $F_{a_i }$.

When system $A$ is composed of MFEs, then the function $F$ can be performed by different distribution of functions between MFEs.

 The logical function $F_A=(x_1,x_2,\dots,x_n)$, which connects the functional states of elements with the states of system operability, is called the system operability function (SMF) \cite{23}.

The matrix of functional resources of considered system is a $(0,1)$ matrix $B(m\times n)=[a_i(f_j)]$ where
\begin{equation*}
    a_i(f_j) = \begin{cases}
    1, & \text{if the employee $a_i$ has the ability to perform function $f_j$},\\
    0, & \text{if MFE $a_i$ does not possess the ability to perform function $f_j$}.
            \end{cases}
\end{equation*}

Then the SMF will be written using the shortest possible ways of successful operation:
\begin{equation}
    S_q = a_{i_1}(f_{j_1})\,\&\,a_{i_2}(f_{j_2})\,\& \cdots \&\,a_{i_m}(f_{j_m})
  \label{eq:5}
\end{equation}
where
\begin{equation*}
  \begin{aligned}
    i_1 &= 1,2,\dots,n-m+1\\
    i_2 &= i_1+1,i_1+2,\dots,n-m+2\\\
    \vdots & \qquad \qquad \vdots\\
    i_m &= i_{m-1} +1, i_{m-1}+2,\dots,n\\
    j_1,j_2,\dots,j_m&\in[1,m], \quad j_1\neq j_2 \neq \cdots \neq j_m, \quad q\in[1,N_s].
  \end{aligned}
\end{equation*}
Here $N_S$ is an indicator of system flexibility, which represents the number of shortest ways for distribution of functions between MFEs and their functioning.

The index of flexibility of the adjustable system $N_S$ is calculated by the formula:
\begin{equation*}
  N_S = \text{per\,B}(m\times n)
\end{equation*}
where $\text{per\,B}$ is a matrix permanent.

It follows from (\ref{eq:5}) that every shortest path includes the
elements equal to 1 of the matrix $B(m\times n)$, which are located in
the intersections of different rows and columns.

It is clear that for $n > m =k_i$, $N_S=\text{per\,B}(m\times n) =
n!/(n-m)!$, and for $n=m=k_i$, $N_S=\text{per\,B}(m\times n)=n!$.

The system operability condition is described by the disjunction of all shortest paths of
functioning:
\begin{equation}
  F_A\big[a_1(f_1),\dots,a_n(f_m)\big] = \bigcup_{q=1}^{N_S} S_q.
  \label{eq:6}
\end{equation}

For example, the shortest ways of functioning of the system composed
of three three-functional MFEs, according to the formula (\ref{eq:5})
are written by the following conjunctions:
\begin{equation*}
  \begin{aligned}
    S_1 & = a_1(f_1) \,\&\, a_2(f_2) \,\&\, a_3(f_3),\qquad &
        S_2 & = a_1(f_1) \,\&\, a_2(f_3) \,\&\, a_3(f_2),\\
    S_3 & = a_1(f_2) \,\&\, a_2(f_1) \,\&\, a_3(f_3), &
        S_4 & = a_1(f_2) \,\&\, a_2(f_3) \,\&\, a_3(f_1),\\
    S_5 & = a_1(f_3) \,\&\, a_2(f_1) \,\&\, a_3(f_2), &
        S_6 & = a_1(f_3) \,\&\, a_2(f_2) \,\&\, a_3(f_1).
  \end{aligned}
\end{equation*}
According to the formula (\ref{eq:6}), the system operability condition composed of three three-functional MFEs can be written in the form of the following disjunction:
\begin{equation}
  F_A\big[a_1(f_i),a_2(f_j),a_3(f_k)\big] = S_1 \cup S_2 \cup S_3 \cup S_4 \cup S_5\cup S_6,
  \label{eq:7}
\end{equation}
where $i\neq j \neq k \in [1,3]$.

The analytical form of describing the performance of a system with a reconfigurable structure composed on the basis of MFEs allows us to obtain formulas for estimating system reliability indicators using logical-probabilistic methods.

When the system is composed of MFEs, then as we mentioned above, in case of partial failure of any of the MFEs in the system, it is possible to carry out such exchange of elements that ensures the continuation of the system's operation. The probability of such a system working without failure is determined by the sum of the probabilities of all operating states:
\begin{align}
  P_A(F)
  & = p_1(f_1,\dots,f_{k_1}) \times p_2(f_1,\dots,f_{k_2}) \times \cdots \times p_m(f_1,\dots,f_{k_m}) \nonumber \\
  &\qquad\qquad + p_1(f_1,\dots,f_{k_1}) \times p_2(f_1,\dots,f_{k_2}) \times \cdots \times p_m(f_1,\dots,f'_{k_m}) + \cdots \nonumber \\
  &\qquad\qquad + p_1(f'_1,\dots,f_{k_1}) \times p_2(f_1,\dots,f_{k_2}) \times \cdots \times p_m(f_1,\dots,f_{k_m}) + \cdots \nonumber \\
  &\qquad\qquad + p_1(f'_1,\dots,f_{k_1}) \times p_2(f_1,f'_2,\dots,f_{k_2}) \times \cdots \times p_m(f_1,\dots,f'_{k_m}) + \cdots \nonumber \\
  &\qquad\qquad + p_1(f'_1,\dots,f_{k_1}) \times p_2(f_1,f'_2,\dots,f_{k_2}) \times \cdots \times p_m(f_1,\dots,f_{k_m}) + \cdots \nonumber \\
  &\qquad\qquad + p_1(f_1,f'_2,\dots,f'_{k_1}) \times p_2(f'_1,f_2,\dots,f'_{k_2}) \times \cdots \times p_m(f'_1,\dots,f'_{k_{m-1}}, f_{k_m}),    \label{eq:8}
\end{align}
where $p_1,p_2,\dots,p_n$ are the probabilities of performing all or part of the set the functions $F=\{f_j|\;j\in[1,m]\}$, by the element $a_i$, $i\in[1,n]$, based on the functional capabilities of each of them $F_{a_i} = \{f_j|\; j\in[1,k_i]\}$.

In the case when $p_1=p_2=\cdots=p_n=p$, the probability of the system working without failure is determined as follows:
\begin{equation}
  P_A(F) = \sum_{\gamma=0}^{k_\Sigma} N_L(\gamma) p^\gamma (1-p)^{k_\Sigma-\gamma},
  \label{eq:9}
\end{equation}
where $N_L(\gamma)$ is a number of working states of the system under conditions of $\gamma$ quantity of missing functions; $p$ are the probabilities of the elements' ability to operate.

\begin{equation}
  N_L(\gamma) = n!\bigg(C_{n^2-n}^{\gamma-n} -\sum_{i=1}^{n!} {\frac{i-1}{i}\,N'_{\gamma_i}}\bigg),\;\;  i\in [1,n!], \;\; \gamma \in [n, n^2].
  \label{eq:10}
\end{equation}
where $N'_{\gamma_i}$ is the number of mutually equivalent variants of the structure, in which the number of ways to implement the function is equal to $i$, and number $n$ of structural elements of failure-free operation is equal to $\gamma$ \cite{20}.

\section{Applied methods}
\label{sec:4}

For the structural analysis of reconfigurable systems and the quantitative assessment of reliability criteria, we employ combinatorial analysis, sampling methods, and logical-probabilistic methods (results are presented in Section \ref{sec:5}).

The results obtained using the sampling method can be found in Tables 1–5 of Section \ref{sec:5}.

Among the logical-probabilistic methods, we used the orthogonalization algorithm, which is based on the formula for the decomposition of Shannon's theorem:
\begin{equation}\label{eq:11-1}
    f(x_1,\dots,x_n)=x_if_1^{(i)}(x_1,\dots,x_n)\vee \overline{x}_if_0^{(i)}(x_1,\dots,x_n).
\end{equation}

A description of the orthogonalization algorithm and the proof of its two main theorems are provided in \cite{17} and \cite{22}.

Using this algorithm, it was possible to transform the logical function of the system operability condition, described in disjunction normal form (DNF), into orthogonal disjunction normal form (ODNF). As is known, in a logical function written by ODNF, each of its members is pairwise orthogonal (their logical product is equal to 0), which allows replacing the logical function with a probabilistic representation. Our goal was to use an algorithm in the application that would allow us to obtain an ODNF with a minimum number of members. For this, we used two theorems, which we offer without proof (the proof of both theorems is given in the book \cite{17}).

\begin{theorem}\label{th:1-1}
The negation $K_i'$ of an  $r$-rank elementary conjunction of $K_i=x_1 x_2\cdots x_r$, which does not contain the negations of logical elements, is equivalent to the following disjunction
\begin{equation}\label{eq:12-1}
    K_i'=x_1'\vee x_1x_2'\vee\cdots\vee x_1x_2\cdots x_{r-1}x_r'.
\end{equation}
Its members are pairwise orthogonal \cite{17}.
\end{theorem}

The \textit{proof} of \eqref{eq:12-1} is straightforward if we apply Shannon's decomposition theorem sequentially for the variables $x_1,x_2,\dots,x_{r-1}$ to the $r$-rank elementary disjunction:
\begin{equation}\label{eq:13-1}
    D_i=x_1^{\alpha_1'}\vee x_2^{\alpha_2'}\vee\cdots\vee x_r^{\alpha_r'},
\end{equation}
which is derived using De Morgan's formula from $K_i=x_1 x_2\cdots x_r$ elementary conjunction.

Indeed, by decomposing the logical function in equation \eqref{eq:13-1} using Shannon's formula with respect to $x_1$ we obtain:
\begin{align}
    x_1^{\alpha_1'}\vee x_2^{\alpha_2'}\vee\cdots\vee x_r^{\alpha_r'} & =
        x_1^{\alpha_1'}\big(1\vee x_2^{\alpha_2'}\vee\cdots\vee x_r^{\alpha_r'}\big)\vee x_1^{\alpha_1}\big(0\vee x_2^{\alpha_2'}\vee\cdots\vee x_r^{\alpha_r'}\big) \nonumber \\
    & =
    x_1^{\alpha_1'}\vee x_1^{\alpha_1}\big(x_2^{\alpha_2'}\vee x_3^{\alpha_3'}\vee\cdots\vee x_r^{\alpha_r'}\big). \label{eq:14-1}
\end{align}

Then, by further decomposing with respect to $x_2,\dots,x_{r-1}$, we obtain the equation \eqref{eq:12-1}.

\begin{theorem}\label{th:2-1}
Boolean logical function $y(x_1,x_2,\dots,x_m)$  represented as DNF
$$  f(x_1,x_2,\dots,x_m)=\bigvee_{i=1}^n K_i(i\leq 2^m),        $$
is equivalent to the function
\begin{equation}\label{eq:15-1}
    f(x_1,x_2,\dots,x_m)=K_1\vee K_1'K_2\vee K_1'K_2'K_3\vee\cdots\vee K_1'K_2'\vee\cdots\vee K_{n-1}'K_n.
\end{equation}
\end{theorem}

The correctness of given transformations is easily proven by De Morgan's formulas and Shannon's decomposition theorem \cite{18}. If instead of each representation $K_i'$ $(i<n)$ we insert its value obtained in the form of \eqref{eq:1}, we get the ODNF of the Boolean function \cite{17}.

The following steps should be performed to convert the Boolean function $y(x_1,\dots,x_m)$ to ODNF:

The validity of the given transformations is easily proven using Shannon's decomposition theorem. If we substitute the value of each expression $K_i'$ $(i\leq n)$ in equation \eqref{eq:15-1} with its value obtained in the form of equation \eqref{eq:12-1}, we obtain the ODNF of the Boolean function $f(x_1,x_2,\dots,x_m)$. Let us describe the algorithm for transforming $y(x_1,x_2,\dots,x_m)$ into the ODNF of a Boolean function:

\begin{enumerate}
\item First, we need to convert the Boolean function $y(x_1,x_2,\dots,x_m)$ into DNF (Disjunctive Normal Form).

\item Next, we renumber the DNF terms from $1$ to $n$ $(n<2m)$ in such a way that lower-rank terms are assigned smaller numbers.
	
\item Using transformation (6.7), we define the ODNF of the function $y(x_1,x_2,\dots,x_m)$.
\end{enumerate}

To reduce the number of operations, we perform the following simplifications in the conjunction $K_1',K_2',\dots,K_{i-1}',K_i'$:  we equate to zero those DNF terms $K_j$ $(j\leq i-1)$, that are orthogonal to the term $K_i'$.

As a result of the transformation of the logical representation of the system operability condition into ODNF, it is possible to replace the logical elements with element-free probabilities and to replace logical operations with arithmetic operations. After this we get the formula for the probability of the system working flawlessly (system reliability):
\begin{equation}\label{eq:16-1}
    P\big\{y(x_1,\dots,x_n)=1\big\}=\sum_{i=1}^s P(R_i),
\end{equation}
where the orthogonal terms of the function are written as ODNF.

To demonstrate the orthogonalization algorithm, let's consider a simple example. Let us consider a 5-element logical function, where the reliability condition is described by the following logical function and diagram:
$$  y(x_1,\dots,x_5)=y(x)=(x_1x_3)V(x_1x_4)V(x_2x_4)V(x_2x_5).     $$

\begin{figure}[h]
\centering
\includegraphics[width=2.7in]   {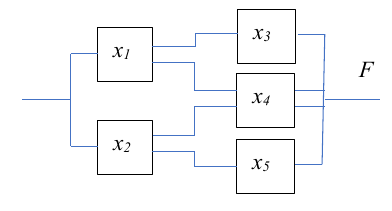}
\caption{}\label{fig:4-1}
\end{figure}

$$  K_1 = x_1 x_3; \quad K_2 = x_1 x_4; \quad K_3 = x_2 x_4; \quad K_3 = x_2 x_5.       $$

Based on Theorem \ref{th:2-1}, we get:
\begin{equation}\label{eq:17-1}
    y(x)=K_1\vee K_1'K_2\vee K_1'K_2'K_3\vee K_1'K_2'K_3'K_4,
\end{equation}
where
$$  K_1'=x_1'\vee x_3'=x_1'\vee x_1x_3'; \quad K_2'=x_1'\vee x_4'=x_1'\vee x_1x_4'; \quad
            K_3'=x_2'\vee x_4'=x_2'\vee x_2x_4'.        $$

By multiplying the conjunctions $K_i$ and their negations, according to formula \eqref{eq:17-1}, we obtain the logical function $y(x)$ written in the ODNF (Optimal Disjunctive Normal Form) as:
$$  y(x)= x_1x_3\vee x_1x_3'x_4\vee x_1'x_2x_4\vee x_1'x_2x_4'x_5\vee x_1x_2x_3'x_4'x_5.    $$

In the resulting ODNF, we can replace the logical variables with probabilistic variables, and logical operations with mathematical operations, which results in:
\begin{multline}\label{eq:18-1}
    P\big\{Y(x)=1\big\}=R_1R_3+R_1R_4(1-R_3)+R_2R_4(1-R_1) \\
    +R_2R_5(1-R_1)(1-R_4)+R_1R_2R_5(1-R_3)(1-R_4),
\end{multline}
where $P$ is the probability of the logical function $Y(x)$ being true, and $R_i$ represents the probabilities of the truth values of logical variables $x_i$, $i\in[1,5]$.

When $R_1 = R_2 = R_3 = R_4 = R_5 = R$, equation \eqref{eq:18-1} takes the form of the following polynomial:
\begin{equation}\label{eq:19-1}
    P\big\{Y(x)=1\big\}=4R^2-3R^3-R^4+R^5.
\end{equation}

By substituting $R=0.5$ into the polynomial \eqref{eq:19-1}, we obtain: $P\{Y(x)=1\}=\delta_A=0.59375$. This corresponds to the reliability condition obtained using the sampling method, where the number of reliable states $(N_{\Omega})=19$ is divided by the total number of possible states $(N_W)=32$:
$$  \delta_A=\frac{N_{\Omega}}{N_W}=\frac{19}{32}=0.59375.      $$

\section{Discussion about structural analysis of multi-core \\ processors}
\label{sec:5}

Considering that for the ratio of structural parameters $n = m > k_i$, the number of all possible states is $N=2^{k_\Sigma}$, and for $n\ge m= k_i$ the number of all possible states is $N=2^{nm}$, of which the number of workable states is more than 50\%, then the number of system
workability states for $n > 8$ is expressed in astronomical numbers. In such a case, to obtain the system reliability estimation polynomial, it becomes necessary to use logical-probabilistic methods - orthogonalization, intersection or recurrent algorithm-based software and to evaluate system efficiency indicators using a computer \cite{17}.

For the structural analysis of multi-core processors, let's make a comparative structural analysis of dual-core processor $A_1$ with structural parameters $n = m = k_i =2$ and of the four-core processor $A_2$ with structural parameters $n = m = k_i= 4$ according to the
system flexibility ($N_S$), reliability ($P_A(F)$), resistance to failure ($g_A(\gamma)$) and structural perfection ($\eta_A$)indicators.

The logical matrix of functional resources of a dual-core processor $n= m = k_i= 2$ has the following form:
\begin{equation*}
  \begin{matrix}
     & \\
    B_A(2\times2) =
  \end{matrix}\!\!\!\!
  \begin{matrix}
    a_1 & a_2 \\
    \left\| \begin{array}{l} 1 \\ 1 \end{array} \right. &
        \left. \begin{array}{r} 1 \\ 1 \end{array} \right\|
  \end{matrix}
\end{equation*}

Based on the formula (\ref{eq:5}), the shortest ways of successful functioning of the system are recorded with the following conjunctions:
\begin{equation*}
  \begin{aligned}
    S_1 &= a_1(f_1) \,\&\, a_2(f_2),\\
    S_2 &= a_1(f_2) \,\&\, a_2(f_1).
  \end{aligned}
\end{equation*}
The system has 2 ways of performing the function $F$, hence its flexibility indicator is $N_S=n!=2$.

If we introduce the notations $a_1(f_1) = x_1$, $a_1(f_2) = x_2$, $a_2(f_1)=x_3$, $a_2(f_2)=x_4$, the shortest ways of successful operation can be written in a form of the following logical matrix (see Table \ref{tab:1}), with the help of which it is possible to calculate the number of system performance states $N_L(\gamma)$, $\gamma\in[0,4]$, and the total number of working states $N_R$.

\begin{table}[htbp]
\begin{center}
\caption{}
\smallskip
  \begin{tabular}{c|cc|cc}\hline\hline
    & & & & \\[-0.3cm]
    $S_q\setminus x_j$ & $x_1$ & $x_2$ & $x_3$ & $x_4$ \\
    & & & & \\[-0.2cm]\hline
    $S_1$ & 1 & 0 & 0 & 1\\[0.05cm]
    $S_2$ & 0 & 1 & 1 & 0\\\hline\hline
  \end{tabular}
  \label{tab:1}
\end{center}
\end{table}

$S_q$ is a way to perform the function $F$, where $q\in[1,n!]$, and $x_j$ is a conditional structural element in the reliability model, $j\in[1,n^2]$.

In Table \ref{tab:2}, for the number of failures $\gamma$, $\gamma\in[1,n^2]$, the numbers of all possible states of the system $N_M(n,\gamma)$, number of operability states $N_L(\gamma)$ and the value of resistance to failure $g_A(\gamma)$ are given. The same table shows the value of $N_R(n) = \sum N_L(\gamma)$ and the structure perfection index $\eta_A$, which is calculated by the formula:
\begin{equation}
  \eta_A = \frac{N_R}{N_\Omega},
  \label{eq:10}
\end{equation}
where $N_\Omega$ is a total number of all possible states, and $N_R$ is the total number of operability states.

\begin{table}[htbp]
\begin{center}
\caption{}
\smallskip
  \begin{tabular}{c|c|c|c}\hline\hline
    & & & \\[-0.3cm]
    $\gamma$ & $N_M(n,\gamma)$ & $N_L(n,\gamma)$ & $g_A(\gamma)$\\
        & & & \\[-0.25cm]\hline
    $0$ & $1$ & $1$ & $1$\\[0.05cm]
    $1$ & $4$ & $2$ & $1$\\[0.05cm]
    $2$ & $6$ & $4$ & $0.333$\\[0.05cm]
    $3$ & $4$ & $0$ & $0$\\[0.05cm]
    $4$ & $1$ & $0$ & $0$\\\hline
    $\Sigma$ & $N_\Omega=16$ & $N_R=7$ & $\eta_A=0.375$ \\\hline\hline
  \end{tabular}
\label{tab:2}
\end{center}
\end{table}

For a dual-core processor, where $N_R(2)=7$, $N_\Omega(2) = 16$, the degree of perfection of the structure will be equal to:
\begin{equation*}
  \eta_A(2) = \frac{N_R(2)}{N_\Omega(2)} = 0.375.
\end{equation*}
To make the system reliability assessment model more understandable, let's consider an example of a system with structural parameters $n =m = k_i = 2$.

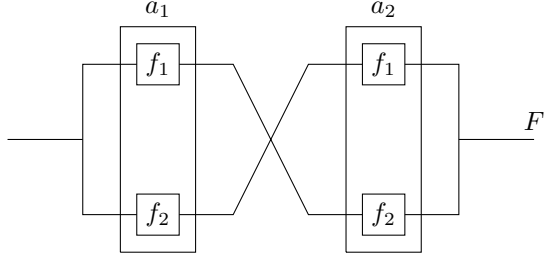
\begin{figure}[htbp]
  \center
  \begin{tikzpicture}
    \draw (-1,0) -- (0,0) (5,0) -- (6,0);
    \draw (0,1) -- ++(0,-2) -- ++(2,0) -- ++(1,2) -- ++(2,0) -- ++(0,-2) -- ++(-2,0) -- ++ (-1,2) -- cycle;
    \draw
    (1,-1) node[rectangle,fill=white,text opacity=1,fill opacity=1,draw] {$f_2$}
    (4,-1) node[rectangle,fill=white,text opacity=1,fill opacity=1,draw] {$f_2$}
    (1,1) node[rectangle,fill=white,text opacity=1,fill opacity=1,draw] {$f_1$}
    (4,1) node[rectangle,fill=white,text opacity=1,fill opacity=1,draw] {$f_1$};
    \draw (1,-1) ++(-.5,-.5)  -- ++(1,0) -- ++(0,3) -- ++(-1,0) -- ++(0,-3);
    \draw (4,-1) ++(-.5,-.5)  -- ++(1,0) -- ++(0,3) -- ++(-1,0) -- ++(0,-3);
    \draw (1,1.5) node[anchor=south] {$a_1$} (4,1.5) node[anchor=south] {$a_2$} (6,0) node[anchor=south] {$F$};
  \end{tikzpicture}
  \caption{$n=m=k_i=2$ system reliability model}\label{fig:4}
\end{figure}

The schematic model of the reliability of a system composed of two dual-functional elements (a dual-core processor is operating in parallel mode), considering (\ref{eq:5}), will have the following form (see Figure \ref{fig:4}).

The general form of the polynomial to calculate the reliability of the system $n = m = k_i = 2$, will be the following
\begin{multline}
  P_A(F)
  = p_1(f_1,f_2) \times p_2(f_1,f_2) \\
  + p_1(f_1,f_2) \times p_2(f_1,f'_2) + p_1(f_1,f_2) \times p_2(f'_1,f_2) + p_1(f_1,f'_2) \times p_2(f_1,f_2) \\
  + p_1(f'_1,f_2) \times p_2(f_1,f_2) + p_1(f_1,f'_2) \times p_2(f'_1,f_2) + p_1(f'_1,f_2) \times p_2(f_1,f'_2).
  \label{eq:11}
\end{multline}

As can be seen from the obtained result, in each set of the formula (\ref{eq:10}) there are different combinations of $(f_1, f_2)$ functions by both MFEs, according to the condition of the system's operability.  For the same values of probabilities of structural elements' operability, using logical-probabilistic methods, a polynomial is obtained:
\begin{equation}
  P_A(F) = 2 p^2 - p^4.
  \label{eq:12}
\end{equation}
When $p=0.99$, the estimation of structural reliability of a dual-core processor will be $P_A(F) = 2 p^2-p^4=0.9996$.

Now let us consider a four-core processor with structural parameters $n = m = k_i = 4$, whose functional resource's matrix has the form:
\begin{equation*}
  B(4\times 4) = \begin{bmatrix}
    1 & 1 & 1 & 1\\
    1 & 1 & 1 & 1\\
    1 & 1 & 1 & 1\\
    1 & 1 & 1 & 1
  \end{bmatrix}
\end{equation*}

The shortest ways for the successful functioning of the system are given by the following
conjunctions:
\allowdisplaybreaks
\begin{gather*}
  \begin{aligned}
    P_F(S_1) &= a_1(f_1) \,\&\, a_2(f_2) \,\&\, a_3(f_3) \,\&\, a_4(f_4), \qquad&
        P_F(S_2) &= a_1(f_1) \,\&\, a_2(f_2) \,\&\, a_3(f_4) \,\&\, a_4(f_3),\\
    P_F(S_3) &= a_1(f_1) \,\&\, a_2(f_3) \,\&\, a_3(f_2) \,\&\, a_4(f_4), &
        P_F(S_4) &= a_1(f_1) \,\&\, a_2(f_3) \,\&\, a_3(f_4) \,\&\, a_4(f_2),\\
    P_F(S_5) &= a_1(f_1) \,\&\, a_2(f_4) \,\&\, a_3(f_2) \,\&\, a_4(f_3),&
        P_F(S_6) &= a_1(f_1) \,\&\, a_2(f_4) \,\&\, a_3(f_3) \,\&\, a_4(f_2),\\
    P_F(S_7) &= a_1(f_2) \,\&\, a_2(f_1) \,\&\, a_3(f_3) \,\&\, a_4(f_4),&
        P_F(S_8) &= a_1(f_2) \,\&\, a_2(f_1) \,\&\, a_3(f_4) \,\&\, a_4(f_3), \\
    P_F(S_9) &= a_1(f_2) \,\&\, a_2(f_3) \,\&\, a_3(f_1) \,\&\, a_4(f_4),&
        P_F(S_{10}) &= a_1(f_2) \,\&\, a_2(f_3) \,\&\, a_3(f_4) \,\&\, a_4(f_1),\\
    P_F(S_{11}) &= a_1(f_2) \,\&\, a_2(f_4) \,\&\, a_3(f_1) \,\&\, a_4(f_3),&
        P_F(S_{12}) &= a_1(f_2) \,\&\, a_2(f_4) \,\&\, a_3(f_3) \,\&\, a_4(f_1), \\
    P_F(S_{13}) &= a_1(f_3) \,\&\, a_2(f_1) \,\&\, a_3(f_2) \,\&\, a_4(f_4),&
        P_F(S_{14}) &= a_1(f_3) \,\&\, a_2(f_1) \,\&\, a_3(f_4) \,\&\, a_4(f_2), \\
    P_F(S_{15}) &= a_1(f_3) \,\&\, a_2(f_2) \,\&\, a_3(f_1) \,\&\, a_4(f_4),&
        P_F(S_{16}) &= a_1(f_3) \,\&\, a_2(f_2) \,\&\, a_3(f_4) \,\&\, a_4(f_1), \\
%      \end{aligned} \\
%\begin{aligned}
    P_F(S_{17}) &= a_1(f_3) \,\&\, a_2(f_4) \,\&\, a_3(f_1) \,\&\, a_4(f_2),&
        P_F(S_{18}) &= a_1(f_3) \,\&\, a_2(f_4) \,\&\, a_3(f_2) \,\&\, a_4(f_1),\\
    P_F(S_{19}) &= a_1(f_4) \,\&\, a_2(f_1) \&y a_3(f_2) \,\&\, a_4(f_3),&
        P_F(S_{20}) &= a_1(f_4) \,\&\, a_2(f_1) \,\&\, a_3(f_3) \,\&\, a_4(f_2),\\
    P_F(S_{21}) &= a_1(f_4) \,\&\, a_2(f_2) \,\&\, a_3(f_1) \,\&\, a_4(f_3),&
        P_F(S_{22}) &= a_1(f_4) \,\&\, a_2(f_2) \,\&\, a_3(f_3) \,\&\, a_4(f_1),\\
    P_F(S_{23}) &= a_1(f_4) \,\&\, a_2(f_3) \,\&\, a_3(f_1) \,\&\, a_4(f_2),&
        P_F(S_{24}) &= a_1(f_4) \,\&\, a_2(f_3) \,\&\, a_3(f_2) \,\&\, a_4(f_1).
 \end{aligned}
\end{gather*}

Binary values $x_i$ of shortest paths for successful system operation are given in Table \ref{tab:3} and $N_L(\gamma)$ and $g_A(\gamma)$ values for different values of $\gamma$ are given in Table \ref{tab:4}.

\begin{table}[htbp]
\begin{center}
\caption{}
\smallskip
  \begin{tabular}{c|cccc|cccc|cccc|cccc}\hline\hline
    & & & & & & & & & & \\[-0.3cm]
    $S_q\setminus x_i$ & $x_1$ & $x_2$ & $x_3$ & $x_4$ & $x_5$ & $x_6$ & $x_7$ & $x_8$ & $x_9$ & $x_{10}$ & $x_{11}$ & $x_{12}$ & $x_{13}$ & $x_{14}$ & $x_{15}$ & $x_{16}$\\
     & & & & & & & & & & \\[-0.2cm]\hline
    $S_{1}$& 1 & 0 & 0 & 0 & 0 & 1 & 0 & 0 & 0 & 0 & 1 & 0 & 0 & 0 & 0 & 1\\[0.05cm]
    $S_{2}$& 1 & 0 & 0 & 0 & 0 & 1 & 0 & 0 & 0 & 0 & 0 & 1 & 0 & 0 & 1 & 0\\[0.05cm]
    $S_{3}$& 1 & 0 & 0 & 0 & 0 & 0 & 1 & 0 & 0 & 1 & 0 & 0 & 0 & 0 & 0 & 1\\[0.05cm]
    $S_{4}$& 1 & 0 & 0 & 0 & 0 & 0 & 0 & 1 & 0 & 1 & 0 & 0 & 0 & 0 & 1 & 0\\[0.05cm]
    $S_{5}$& 1 & 0 & 0 & 0 & 0 & 0 & 1 & 0 & 0 & 0 & 0 & 1 & 0 & 1 & 0 & 0\\[0.05cm]
    $S_{6}$& 1 & 0 & 0 & 0 & 0 & 0 & 0 & 1 & 0 & 0 & 1 & 0 & 0 & 1 & 0 & 0\\[0.05cm]
    $S_{7}$& 0 & 1 & 0 & 0 & 1 & 0 & 0 & 0 & 0 & 0 & 1 & 0 & 0 & 0 & 0 & 1\\[0.05cm]
    $S_{8}$& 0 & 1 & 0 & 0 & 1 & 0 & 0 & 0 & 0 & 0 & 0 & 0 & 0 & 0 & 1 & 0\\[0.05cm]
    $S_{9}$& 0 & 0 & 1 & 0 & 1 & 0 & 0 & 0 & 0 & 1 & 0 & 0 & 0 & 0 & 0 & 1\\[0.05cm]
    $S_{10}$& 0 & 0 & 0 & 1 & 1 & 0 & 0 & 0 & 0 & 1 & 0 & 0 & 0 & 0 & 1 & 0\\[0.05cm]
    $S_{11}$& 0 & 0 & 1 & 0 & 1 & 0 & 0 & 0 & 0 & 0 & 0 & 1 & 0 & 1 & 0 & 0\\[0.05cm]
    $S_{12}$& 0 & 0 & 0 & 1 & 1 & 0 & 0 & 0 & 0 & 0 & 1 & 0 & 0 & 1 & 0 & 0\\[0.05cm]
    $S_{13}$& 0 & 1 & 0 & 0 & 0 & 0 & 1 & 0 & 1 & 0 & 0 & 0 & 0 & 0 & 0 & 1\\[0.05cm]
    $S_{14}$& 0 & 1 & 0 & 0 & 0 & 0 & 0 & 1 & 1 & 0 & 0 & 0 & 0 & 0 & 1 & 0\\[0.05cm]
    $S_{15}$& 0 & 0 & 1 & 0 & 0 & 1 & 0 & 0 & 1 & 0 & 0 & 0 & 0 & 0 & 0 & 1\\[0.05cm]
    $S_{16}$& 0 & 0 & 0 & 1 & 0 & 1 & 0 & 0 & 1 & 0 & 0 & 0 & 0 & 0 & 1 & 0\\[0.05cm]
    $S_{17}$& 0 & 0 & 1 & 0 & 0 & 0 & 0 & 1 & 1 & 0 & 0 & 0 & 0 & 1 & 0 & 0\\[0.05cm]
    $S_{18}$& 0 & 0 & 0 & 1 & 0 & 0 & 1 & 0 & 1 & 0 & 0 & 0 & 0 & 1 & 0 & 0\\[0.05cm]
    $S_{19}$& 0 & 1 & 0 & 0 & 0 & 0 & 1 & 0 & 0 & 0 & 0 & 1 & 1 & 0 & 0 & 0\\[0.05cm]
    $S_{20}$& 0 & 1 & 0 & 0 & 0 & 0 & 0 & 1 & 0 & 0 & 1 & 0 & 1 & 0 & 0 & 0\\[0.05cm]
    $S_{21}$& 0 & 0 & 1 & 0 & 0 & 1 & 0 & 0 & 0 & 0 & 0 & 1 & 1 & 0 & 0 & 0\\[0.05cm]
    $S_{22}$& 0 & 0 & 0 & 1 & 0 & 1 & 0 & 0 & 0 & 0 & 1 & 0 & 1 & 0 & 0 & 0\\[0.05cm]
    $S_{23}$& 0 & 0 & 1 & 0 & 0 & 0 & 0 & 1 & 0 & 1 & 0 & 0 & 1 & 0 & 0 & 0\\[0.05cm]
    $S_{24}$& 0 & 0 & 0 & 1 & 0 & 0 & 1 & 0 & 0 & 1 & 0 & 0 & 1 & 0 & 0 & 0\\\hline\hline
  \end{tabular}
\label{tab:3}
\end{center}
\end{table}

\begin{table}[h]
\begin{center}
\caption{}
\smallskip
  \begin{tabular}{c|c|c|c}\hline\hline
    & & & \\[-0.3cm]
    $\gamma$ & $N_M(n,\gamma)$ & $N_L(n,\gamma)$ & $g_A(\gamma)$\\
    & & & \\[-0.2cm]\hline
    0 & 1 & 1 & 1\\[0.05cm]
    1 & 16 & 16 & 1\\[0.05cm]
    2 & 120 & 120 & 1\\[0.05cm]
    3 & 560 & 560 & 1\\[0.05cm]
    4 & 1820 & 1812 & 0.996\\[0.05cm]
    5 & 4368 & 4272 & 0.978\\[0.05cm]
    6 & 8008 & 7432 & 0.928\\[0.05cm]
    7 & 11440 & 9312 & 0.814\\[0.05cm]
    8 & 12870 & 8010 & 0.622\\[0.05cm]
    9 & 11440 & 4464 & 0.390\\[0.05cm]
    10 & 8008 & 1512 & 0.189\\[0.05cm]
    11 & 4368 & 288 & 0.0659\\[0.05cm]
    12 & 1820 & 24 & 0.0132\\[0.05cm]
    13 & 560 & 0 & 0\\[0.05cm]
    14 & 120 & 0 & 0\\[0.05cm]
    15 & 16 & 0 & 0\\[0.05cm]
    16 & 1 & 0 & 0\\\hline
    $\Sigma$ & $N_\Omega = 65536$ & $N_R=37823$ & $\eta_A=0.57713$\\\hline\hline
  \end{tabular}
  \label{tab:4}
  \end{center}
\end{table}

Since $N_\Omega(4)=2^{16}=65536$ and $N_R(4)=37823$, the structure perfection indicator calculated by formula (\ref{eq:10}) will be equal to $\eta_A(4)=0.57713$.

For a quad-core processor $n = m = k_i=4$, for the same values of the probabilities of the structural elements' operability, using logical-probabilistic methods the following polynomial for the estimation of system reliability is obtained:
\allowdisplaybreaks[0]
\begin{multline*}
  P_A(F)=  24 p^4 -  72 p^6 -  96 p^7 +  234 p^8 +  528 p^9 -  1808 p^{10} \\
  + 2160 p^{11} -  1392 p^{12} +  528 p^{13} -  120 p^{14} +  16 p^{15} -  p^{16}.
\end{multline*}
When $p=0.99$, $P_A(F)=0.9999991$, which indicates the higher reliability of a 4-core processor compared to a 2-core processor.

If we consider the example, then a 2-core processor with 4 operational blocks, $n=2$, $m=4$, can be described by the following $(0,1)$ matrix:
$$  B_A\big[a_i(f_j)\big]=\begin{Vmatrix}
                a_1 (f_1 ) & a_1 (f_2 ) & a_1 (f_3 ) & a_1 (f_4 ) \\
                a_2 (f_1 ) & a_2 (f_2 ) & a_2 (f_3 ) & a_2 (f_4 )
                        \end{Vmatrix},      $$
where $a_i (f_j )=1$ if $j$ block of $a_i$ core performs $f_j$ function and $a_i (f_j )=0$, in opposite case.

If we introduce notations:
\begin{gather*}
    a_1 (f_1 ) =x_1, \quad a_1 (f_2 )=x_2, \quad a_1 (f_3 )=x_3, \quad a_1 (f_4 ) =x_4, \\
    a_2 (f_1 )=x_5, \quad a_2 (f_2 )=x_6, \quad a_2 (f_3 ) =x_7, \quad a_2 (f_4 )=x_8,
\end{gather*}
then the shortest paths for the operation of a 2-core processor are written by the following conjunctions:
\begin{gather*}
    S_1=x_1 x_2 x_3 x_4, \quad S_2=x_1 x_2 x_3 x_8, \quad S_3=x_1 x_2 x_4 x_7, \quad S_4=x_1 x_3 x_4 x_6, \\
    S_5=x_2 x_3 x_4 x_5,  \quad S_6=x_1 x_2 x_7 x_8,  \quad S_7=x_1 x_3 x_6 x_8,  \quad S_8=x_2 x_3 x_5 x_8, \\
    S_9=x_1 x_4 x_6 x_7,  \quad S_{10}=x_2 x_4 x_5 x_7, \quad  S_{11}=x_3 x_4 x_5 x_6,  \quad   S_{12}=x_1 x_6 x_7 x_8, \\
    S_{13}=x_2 x_5 x_7 x_8,  \quad S_{14}=x_3 x_5 x_6 x_8,  \quad S_{15}=x_4 x_5 x_6 x_7,  \quad S_{16}=x_5 x_6 x_7 x_8.
\end{gather*}
The same ways of functioning with $(0,1)$ logical variables are given in Table~\ref{table:5-1}.

\begin{table}[h]
\begin{center}
\caption{}
\smallskip
\begin{tabular}{p{0.7cm}|p{0.7cm}|p{0.7cm}|p{0.7cm}|p{0.7cm}|p{0.7cm}|p{0.7cm}|p{0.7cm}|p{0.7cm}}
\hline\hline%\hline\noalign{\smallskip}
     & \multicolumn{4}{c}{$a_1$} & \multicolumn{4}{|c}{$a_2$} \\\hline
%\noalign{\smallskip}\svhline\noalign{\smallskip}
    $A/F$ & $f_1$ & $f_2$ & $f_3$ & $f_4$ & $f_5$ & $f_6$ & $f_7$ & $f_8$ \\\hline
    $X$ & $x_1$ & $x_2$ & $x_3$ & $x_4$ & $x_5$ & $x_6$ & $x_7$ & $x_8$ \\\hline
    1 & 1 & 1 & 1 & 1 & 0 & 0 & 0 & 0 \\[0.05cm]
    2 & 1 & 1 & 1 & 0 & 0 & 0 & 0 & 1 \\[0.05cm]
    3 & 1 & 1 & 0 & 1 & 0 & 0 & 1 & 0 \\[0.05cm]
    4 & 1 & 0 & 1 & 1 & 0 & 1 & 0 & 0 \\[0.05cm]
    5 & 0 & 1 & 1 & 1 & 1 & 0 & 0 & 0 \\[0.05cm]
    6 & 1 & 1 & 0 & 0 & 0 & 0 & 1 & 1 \\[0.05cm]
    7 & 1 & 0 & 1 & 0 & 0 & 1 & 0 & 1 \\[0.05cm]
    8 & 0 & 1 & 1 & 0 & 1 & 0 & 0 & 1 \\[0.05cm]
    9 & 1 & 0 & 0 & 1 & 0 & 1 & 1 & 0 \\[0.05cm]
    10 & 0 & 1 & 0 & 1 & 1 & 0 & 1 & 0 \\[0.05cm]
    11 & 0 & 0 & 1 & 1 & 1 & 1 & 0 & 0 \\[0.05cm]
    12 & 1 & 0 & 0 & 0 & 0 & 1 & 1 & 1 \\[0.05cm]
    13 & 0 & 1 & 0 & 0 & 1 & 0 & 1 & 1 \\[0.05cm]
    14 & 0 & 0 & 1 & 0 & 1 & 1 & 0 & 1 \\[0.05cm]
    15 & 0 & 0 & 0 & 1 & 1 & 1 & 1 & 0 \\[0.05cm]
    16 & 0 & 0 & 0 & 0 & 1 & 1 & 1 & 1 \\
\hline\hline%\noalign{\smallskip}\hline\noalign{\smallskip}
\end{tabular}
\label{table:5-1}
\end{center}
\end{table}

Using our application based on combinatorial analysis and logical-probabilistic methods, the following results were obtained:
\begin{itemize}
\item When $p_i=0.99$, where $p_i$ -- is the probability of the functionality of each functional block, the probability of the processor's faultlessness is $P(A_F )=0.9996$;

\item The number of all possible states -- $N_{\Omega}=2^nm=2^8=256$;

\item The number of operational states -- $N_R=81$;
	
\item The coefficient of structural perfection of the processor -- $\eta_A=N_R/N_{\Omega}=0.31640625$.
\end{itemize}

Thus, in Section \ref{sec:5} of the article, we discussed two cases:
\begin{enumerate}
\item The combinatorial structural analysis of 2-core and 4-core processors, with evaluations of flexibility, reliability, and fault tolerance;

\item The combinatorial structural analysis of the functional blocks of a 2-core processor, with modeling of the interchanging functions between blocks and evaluations of flexibility, reliability, and fault tolerance.
\end{enumerate}

\section{Conclusion}
\label{sec:6}

In the paper, we discuss methods for structural analysis and quantitative evaluation of reliability metrics for multi-core systems.

Models for evaluating the reliability indicators of multi-core systems have been developed, taking into account the following limitations: Only cases where cores or blocks execute tasks or functions in parallel mode are considered; the cases of independent and non-recoverable partial failures of system elements are also addressed. Additionally, in the reliability model of the multi-core system, partial failures of elements are handled by considering immediate switches to operational paths that ensure fault tolerance and the continued successful operation of the system.

The structural analysis and evaluation of efficiency indicators (reliability, fault tolerance, viability, flexibility) of multi-core processor composed with multi-functional cores have been obtained. Using logical-probabilistic methods, the following has been developed:
\begin{itemize}
\item Models for evaluating the reliability and resistance to failure   of processor cores as multifunctional elements;

\item Logical-probabilistic models of the shortest paths, flexibility and performance conditions for successful operation of multi-core processors based on multi-functional cores;

\item Models for estimating reliability, resistance to failure and viability of multi-core processors considering all possible states of performance.
\end{itemize}

Based on the comparative analysis of the structural reliability indicators of two-core and four-core processors, it was concluded that the more cores in a processor, the higher the flexibility, structural perfection, reliability and fault tolerance indicators, and therefore, the more viable a multi-core processor is.

%\bibliographystyle{unsrt}
%\bibliography{criad001}

\section*{Conflict of interests}

The authors declare no conflict of interests.

\end{document}